\documentclass[aps,pra,twocolumn,floatfix,superscriptaddress]{revtex4}  
\usepackage{float}
\usepackage{graphicx}
\usepackage{amssymb}
\usepackage{dsfont}
\usepackage{amsmath}  
\usepackage{epstopdf}
\usepackage{xcolor}
\begin{document}

\def\ket#1{|#1\rangle} 
\def\bra#1{\langle#1|}
\def\av#1{\langle#1\rangle}
\def\dkp#1{\kappa+i(\Delta+#1)}
\def\dkm#1{\kappa-i(\Delta+#1)}
\def\pp{{\prime\prime}}
\def\ppp{{\prime\prime\prime}}
\def\w{\omega}
\def\k{\kappa}
\def\D{\Delta}
\def\wp{\omega^\prime}
\def\wpp{\omega^{\prime\prime}}

\def\ap{a^{\prime}}
\def\ad{a^{\dagger}}

\title{Rotation Errors Due to Field Quantization for Simultaneously Driven Atoms}
\author{Hunter Lindemann}
\affiliation{University of Arkansas, AR, USA}
\author{Shanon Vuglar}
\affiliation{John Brown University, AR, USA}
\author{Julio Gea-Banacloche}
\affiliation{University of Arkansas, AR, USA}

\date{\today}

\begin{abstract}
When an electromagnetic field in a coherent or quasiclassical (e.g., squeezed) state is used to simultaneously drive an ensemble of two-level atoms, the quantum nature of the field will, in general, cause the final state of the atoms to differ from the one predicted for a totally classical field.  This is a potential source of error in quantum logic gates in which the gate is the rotation of the atoms by a laser. In this paper, we use second order perturbation theory to find how this error scales with the number of atoms, $N$, being driven simultaneously, for an arbitrary rotation angle.  The result depends on the initial atomic state: for some highly entangled states, and a field in a coherent state, the error may scale as $N^2$, yet we find that the average over a random distribution of initial states only scales as $N$.  We discuss possible ways to mitigate the error, including the use of squeezed states, as well as adjusting the interaction time between the field and atoms to be different from what would be expected  from the classical-field treatment.
\end{abstract}
\maketitle

\section{Introduction}
It has been known for some time \cite{ozawa,jgbpra,jgbprl}  that the quantum nature of the  electromagnetic field can be a source of error in the performance of logical gates carried out on qubits driven by such fields (such as two-level atoms or ions, or, in principle, even superconducting qubits).  Although this is generally a very small error, scaling as $1/\bar n$ (where $\bar n$ is the average number of photons in the pulse used to manipulate the two-level system), it does set a lower limit on the power needed to perform logical gates such as single-qubit rotations (and potentially more complicated gates as well); hence, this is relevant in the context of the recently-proposed ``quantum energy initiative'' \cite{auffeves}.

A less well-known result \cite{jgbozawa} is that, in some cases, when one tries to use the same pulse to carry out an operation on $N$ qubits, the total error may actually scale as $N^2$: that is, the error per qubit is larger than if independent pulses were used.  For sequential operations (one qubit at a time), this means that it would be, in general, detrimental to reuse the same control pulse.  This has led to an ingenious proposal \cite{ikonen} of a passive scheme, involving ancillary qubits, to ``restore'' the pulse so it can be reused many times, while keeping the error per qubit close to what would be obtained with a ``fresh'' pulse.  In a recent paper \cite{shanon}, we have shed light on how this ancilla-based ``restoration'' process works, and unveiled an interesting parallel with the so-called ``transcoherent'' states \cite{transc1,transc2}.

We note that the bounds derived in \cite{jgbozawa} apply, in principle, regardless of whether the interaction of the field pulse with the $N$ qubits is sequential or simultaneous.  In the latter case, of course, the scheme of \cite{ikonen} is not applicable.  There are, in fact, a number of situations, arising in the context of quantum computation, where one might want to use a single pulse to manipulate a large set of qubits all at once: these include initialization of the working qubits, as well as error correction with transversal gates \cite{knill,brown}.  

Our goal in this paper is to study the errors caused by the quantum nature of the field under the simple assumption that the field is used to carry out a rotation of $N$ qubits simultaneously.  We study the error dependence on the number of atoms (and photons), the angle of rotation, the statistics of the field (coherent or squeezed), and the interaction time, allowing the latter to depart slightly from the nominal time expected from a classical-field treatment.  We obtain analytical results for some of the ``worst-case'' initial atomic states, as well as for the average of the error over a totally random initial state, and compare them to numerical simulations over random ensembles.  The latter allow us to go beyond the limitations of the second-order perturbation theory used in our analytical treatment, and also to visualize what might be the ``typical''  spread of the errors around the average in a possible implementation.

The paper is structured as follows: in Section 2 we introduce our model and methods.  In Section 3 we derive the error for initial entangled atomic states of the ``cat'' type, which yield the worst-case scenario ($N^2$) error scaling, and show how this can be reduced to an $N\sqrt N$ scaling by using an appropriate squeezed state instead of a coherent state.  In Section 4 we derive the result for the average error, find the squeezed states that minimize it (which are different from the ones in Section 3), and compare with the results from numerical simulations.  Finally, in Section 5 we explore the possibility of reducing the error by changing the interaction time (a possibility that was suggested by some of our previous numerical studies).  Section 6 briefly summarizes our results and their possible significance.

\section{Model and procedure}
 We begin with the interaction picture Hamiltonian of $N$ atoms being driven by a single-mode electromagnetic field. We assume everything is at resonance and make the rotating wave approximation,
\begin{equation}
	H=g(J_+a+J_-\ad)
	\label{eq: 1}
\end{equation}
where $g$ is the single atom-field coupling constant, and the angular momentum-like operators $J_+, J_-$ are built from the raising and lowering operators for the individual atoms in the standard way: $J_+ = \hbar \sum_{i=1}^N \ket e_i \bra g$, $J_- = J_+^\dagger$, with $\ket e$ and $\ket g$ being the excited and ground states, respectively.  

If the driving field is quasiclassical,  (e.g., coherent or squeezed), meaning it has a relatively large coherent amplitude $\alpha \equiv \av{a}$ compared to its phase-space uncertainty, we can define $a^\prime = a -\alpha$ and write
\begin{align}
a &= \alpha+\ap \cr
\ad &=  \alpha^\ast+{\ap}^\dagger
\label{e2}
\end{align}
where the idea is to treat $\alpha^\prime$ subsequently as a small perturbation.  We will take $\alpha$ to be real, in what follows, without loss of generality; this means the quadrature operator $\ap+{\ap}^\dagger$ will be associated with amplitude fluctuations and $\ap-{\ap}^\dagger$ with phase fluctuations.  We expect $|\alpha|^2$ to be of the order of magnitude of the (average) number of photons in the field, $\bar n$, which we take to be relatively large.   This breaks the Hamiltonian into a ``classical field'' part, $H_0=2g \alpha J_x$, and a ``quantum field'' part.  The classical field, acting for a time $t$, will produce a collective rotation (around the $x$ axis) by an angle  $\theta = 2g\alpha t$, so in the first few sections we will take the total interaction time $t$ to be equal to $\theta/2 g\alpha$, if $\theta$ is the desired rotation angle; in Section 5 we will explore what happens if $t$ deviates from this value by a small amount $\delta t$.  
 
 We proceed by going to a new interaction picture, $\ket{\Psi}_I = e^{i H_0 t/\hbar}\ket{\Psi}$, where we eliminate the classical field part $H_0$, and are left only with the time-dependent Hamiltonian $H_I = e^{i H_0 t/\hbar} H e^{-i H_0 t/\hbar}$ given by
\begin{align}
	H_I = &gJ_x(\ap+{\ap}^\dagger)\cr
	&+g\left(i\cos({2g\alpha t})J_y -i\sin({2g\alpha t})J_z\right)(\ap-{\ap}^\dagger) 
\label{eq: 2}
\end{align}
which we will treat perturbatively.  Since the total evolution time scales as $\sim 1/\sqrt{\bar n}$, we find that each order of perturbation adds a factor of $1/\sqrt{\bar{n}}$ to the final result. To achieve the desired order of $1/\bar{n}$, we will need to apply second order perturbation. 

Let the initial atomic state be $\ket{\psi_0}$.  Our nominal target state, in the original picture, is this state rotated by $\theta$, i.e., $e^{-i J_x \theta/\hbar} \ket{\psi_0}$; in the new interaction picture, this just becomes $\ket{\psi_0}$.  Hence, to get the fidelity, we project the perturbative state onto $\ket{\psi_0}$.  The result can be written as 
\begin{equation}
\av{\psi_0|\Psi_0} + \av{\psi_0|\Psi^{(1)}} + \av{\psi_0|\Psi^{(2)}} 
\label{e4}
\end{equation}
where $\ket{\Psi_0} = \ket{\psi_0}\ket{\Phi_0}$ is the initial joint atom-field state (so $\ket{\Phi_0} = \av{\psi_0|\Psi_0}$ is the initial field state) and 
 \begin{align}
 \ket{\Psi^{(1)}(t)} &=-\frac{i}{\hbar} \int_0^t H_I(t^\prime)\ket{\Psi_0}\, dt^\prime \cr
 \ket{\Psi^{(2)}(t)} &=-\frac{i}{\hbar} \int_0^t H_I(t^\prime)\ket{\Psi^{(1)}(t^\prime)}\, dt^\prime
 \label{e6}
\end{align}
Equation (\ref{e4}) is the field state associated with the outcome $\ket{\psi_0}$ for the atomic ensemble.  The norm squared of that field state gives the probability of the corresponding outcome.  Hence, neglecting terms of order higher than $1/\bar n$, the error probability can be written as 
\begin{align}
\epsilon &= 1- \| \ket{\Phi_0}+ \av{\psi_0|\Psi^{(1)}} + \av{\psi_0|\Psi^{(2)}}\|^2 \cr
&=-2Re\av{\Psi_0 | \Psi^{(1)}}-\|\av{\psi_0|\Psi^{(1)}}\|^2-2Re\av{\Psi_0|\Psi^{(2)}}
	\label{eq: 3}
 \end{align}
We note that, since we have separated out the coherent amplitude $\alpha$ in Eq.~(\ref{e2}),  the first term on the right-hand side of (\ref{eq: 3}) will vanish,  as $\av{(\ap+{\ap}^\dagger)} = \av{(\ap-{\ap}^\dagger)} = 0$ over the initial field state.  The last two terms go as $1/\bar{n}$, and they are the only ones that contribute to the error to this order.  In the following, we will use Eq.~(\ref{eq: 3}) to derive general expressions valid for an arbitrary number of atoms $N$, noting only that the $N=1$ and $N=2$ cases need to be handled separately.

\section{Initial Cat State}
\subsection{$N>2$ Case}
In Ref.~\cite{jgbozawa} it was argued that an error scaling proportional to $N^2$ would be obtained for cat states of the form
\begin{equation}
\ket{\psi_z} = \frac{1}{\sqrt 2} \bigl((\ket{ggg...} + \ket{eee...}\bigr)
\label{catz}
\end{equation}
and that this error was directly related to phase fluctuations.  Since we are working with (pseudo-)angular momentum operators, it is easier to consider this state in the angular momentum basis of eigenstates of $J_z$, labeled by $\ket{j,m}$:
\begin{equation}
\ket{\psi_z} = \frac{1}{\sqrt{2}}\left(\biggl|\frac{N}{2},-\frac{N}{2}\biggr\rangle+\biggl|\frac{N}{2},\frac{N}{2}\biggr\rangle\right)
\end{equation}
(since the effective angular momentum of this subspace is $j=N/2$), which accounts for the subscript $z$ on $\ket{\psi_z}$.  Later on in this section, we will consider as well the corresponding cat states $\ket{\psi_x}$, formed out of maximal angular momentum eigenstates of the operator $J_x$.

It is easy to see that in the state (\ref{catz}) the expectation values $\av{J_x}$, $\av{J_y}$ and $\av{J_z}$ all vanish, and hence the $\av{\Psi_0|\Psi^{(1)}}$ term in (\ref{eq: 3}) does not contribute, which leaves only the $-2Re\av{\Psi_0|\Psi^{(2)}}$ term to consider.  If we write the $J_x$ and $J_y$ operators in terms of raising and lowering operators,  we see that, for $N=3$ and larger, the only products that reproduce the initial state are $J_+J_-$, $J_-J_+$, or $J_zJ_z$.  The projection of the second order perturbation onto the initial state then gives
\begin{align}
	\av{\Psi_0 | \Psi^{(2)}} =  &-\frac{ e^{-2r}}{8 \alpha^2\hbar^2}\theta^2 \av{J_x^2}\cr
	&-\frac{e^{2r}}{8 \alpha^2\hbar^2}\left(\av{J_y^2}\sin^2\theta +4\av{J_z^2}\sin^4\frac\theta 2 \right)
	\label{eq: 4}
\end{align}
where we have integrated Eqs.~(\ref{e6}) up to the time $t=\theta/2g\alpha$.  The first term in Eq.~(\ref{eq: 4}) is the field amplitude error, arising from $\av{(\ap+{\ap}^\dagger)^2}$.  The remaining terms give the phase quadrature error.  We have allowed for the possibility of using a squeezed state, with squeezing parameter $r$, to possibly reduce one term at the expense of the other, with a positive value of $r$ corresponding to amplitude squeezing.  

We observe that, in the $z$-cat state (\ref{catz}), one has $\av{J_x^2} = \av{J_y^2} = \hbar^2 N/4$, whereas $\av{J_z^2} = \hbar^2 N^2/4$; it is the latter that is responsible for the  $N^2$ scaling of the error, arising in this case from the phase fluctuations term (in agreement with the predictions in \cite{jgbozawa}).  Somewhat surprisingly, though, it is not the $z$-cat state that yields the largest error in the absence of squeezing.  Indeed, since the coefficient of the amplitude fluctuations, $\theta^2$, is larger, for all $0\le\theta\le\pi$, than the phase fluctuation term $4\sin^4(\theta/2)$, one could get an even larger error by starting from an $x$-cat state, for which $\av{J_y^2} = \av{J_z^2} = \hbar^2 N/4$ but $\av{J_x^2} = \hbar^2 N^2/4$.  We find, then, for each of these two cases, a total error  
given by
\begin{align}
\epsilon_\text{$z$-cat} &= \frac{N}{16 \alpha^2}\left[\theta^2e^{-2r}+\left(\sin^2\theta +4N\sin^4\frac\theta 2 \right)e^{2r}\right] \cr
\epsilon_\text{$x$-cat} &= \frac{N}{16 \alpha^2}\left[N\theta^2e^{-2r}+4\sin^2\frac\theta 2\,e^{2r}\right]
\label{e10}
\end{align}

For reference, explicitly, the $x$-cat state reads
\begin{equation}
\ket{\psi_x} = \frac{1}{\sqrt{2}}\left(\ket{+++\ldots} + \ket{---\ldots}\right)
\end{equation}
with $\ket \pm = (\ket e \pm \ket g)/\sqrt 2$.  There is some irony here, since the $x$-cat state should be left invariant by a rotation around the $x$ axis (and would be, for a classical field); instead, it shows some of the largest possible error when a quantum field is used!

The results (\ref{e10}) are plotted in Fig.~1 for $N=4$, both for zero squeezing and for the optimal squeezing, which is obtained by minimizing the expressions (\ref{e10}) with respect to $r$.  The result is
\begin{align}
r_\text{opt,$z$} &= \frac{1}{4}\ln\left(\frac{\theta^2}{\sin^2{\theta}+4N\sin^4(\theta/2)}\right) \cr
r_\text{opt,$x$} &= \frac{1}{4}\ln\left(\frac{N\theta^2}{4\sin^2(\theta/2)}\right)
	\label{eq: 8}
\end{align}
For large enough $N$, $r_{opt,z}$ is negative, scaling as $\ln(N^{-1/4})$ (corresponding to squeezing of the phase quadrature), while  $r_{opt,x}$ is positive, scaling as $\ln(N^{1/4})$  (corresponding to squeezing of the amplitude quadrature).

\begin{figure}
\includegraphics[width=8cm]{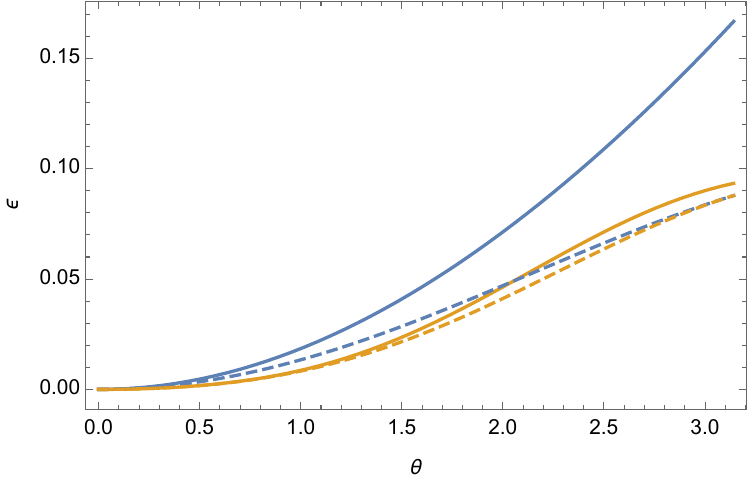}
\caption{The error for an $x$-cat state (upper curve), and for a $z$-cat state (lower curve), for $r=0$ (coherent state, solid lines) and for the corresponding optimally squeezed state (dashed lines), for $5$ atoms and approximately $100$ photons ($\alpha =10$), as a function of the angle of rotation $\theta$.}
\end{figure}

The figure shows that the error for the $x$-cat state with an unsqueezed (coherent) field is, typically, substantially larger than that for the $z$-cat state, but can be dramatically reduced by using the appropriate squeezed field. It is worth noting that squeezing the field increases the average photon count, so that $\bar{n}=|\alpha|^2+\sinh^2r$ where the first term represents the coherent part of the field and the latter represents the contribution from squeezing.   Nevertheless, it follows from Eq.~(\ref{eq: 8}) that, for large $N$, $\sinh^2r = \frac 1 4 (e^{r} - e^{-r})^2$ scales only as $\sqrt N$ (for both the $x$ and $z$ cat states), which is presumably much smaller than $\bar n$ in the first place.  Figure 2 shows the required extra photons for optimal squeezing in the case of an $x$-cat state, for $N=5, 10$, and $20$ atoms.

 \begin{figure}
\includegraphics[width=8cm]{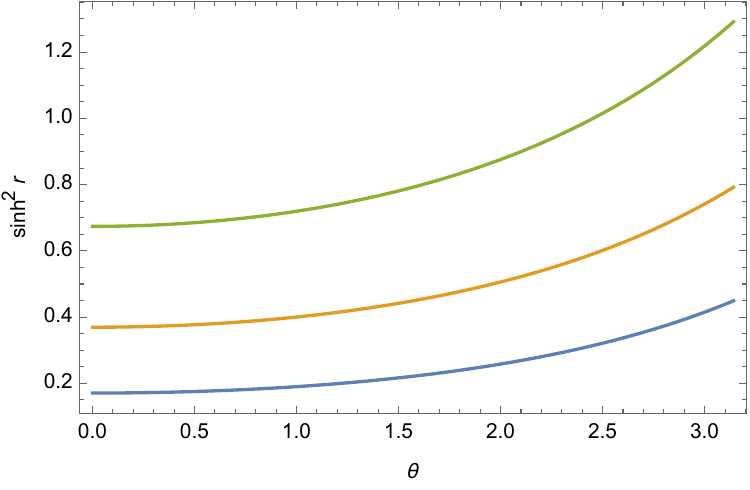}
\caption{The number of photons added by the optimal squeezing of a field to rotate an $x$-cat state by an angle $\theta$, for $N=5, 10$, and $20$ atoms (bottom to top)}
\end{figure}

What one gains from squeezing the field, in either case, is a much more favorable scaling of the maximum error, as $N\sqrt N$ instead of $N^2$.  Explicitly, we have
\begin{align}
\epsilon_\text{opt,$z$} &= \frac{N}{8 \alpha^2}\left[\theta\sqrt{\sin^2\theta +4N\sin^4\frac\theta 2 }\right] \cr
\epsilon_\text{opt,$x$} &= \frac{N\sqrt N}{4 \alpha^2}\theta\sin\frac\theta 2 
\label{e13}
\end{align}
with both yielding a maximum error of $N\sqrt N\pi/4\alpha^2$ for $\theta=\pi$.

\subsection{$N=1,2$ Cases}
As stated before, N=1 and N=2 each produce their own unique error equations with corresponding squeezing parameter. For two atoms, the projection of the first order perturbation is still zero as before, but now we find  $\av{J_x^2} = \av{J_z^2} =\hbar^2$ and $\av{J_y^2} =0$.  In fact, for this case the $x$- and $z$-cat states coincide.  The resulting error is

\begin{equation}
\epsilon_{2}=\frac{1}{4\alpha^2}\left(\theta^2 e^{-2r}+4 \sin^4\frac\theta 2 e^{2r}\right)
	\label{eq: 11}
 \end{equation}
This error is minimized for the squeezing parameter given by
\begin{equation}
r=\frac{1}{4}\ln\left(\frac{\theta^2}{4 \sin^4(\theta/2)}\right)
	\label{eq: 12}
\end{equation}
which produces the optimized error of 
\begin{equation}
\epsilon_\text{2,opt}=\frac{1}{\alpha^2}\theta\sin^2\frac \theta 2
	\label{eq: 13}
\end{equation}
Finally, for completeness, we consider the $N=1$ case.  This is just the Jaynes-Cummings model, and finding the initial state leading to the maximal error (always to order $1/\alpha^2$) is relatively straightforward.  Somewhat surprisingly, it turns out to be the excited state $\ket e$.  The result is
\begin{equation}
\epsilon_1 = \frac{1}{16\alpha^2}\left(\theta e^{-r} +\sin\theta e^r \right)^2
\label{e17}
\end{equation}
The result for an atom initially in the ground state is like above but with the two terms in parenthesis being subtracted instead of added.  The difference seems to be due to the possibility of spontaneous emission from the excited state.  Optimal squeezing requires 
\begin{equation}
r = \frac 1 2 \ln\left(\frac{\theta}{\sin\theta}\right)
\end{equation}
leading to 
\begin{equation}
\epsilon_\text{1,opt} =  \frac{1}{16\alpha^2}\theta\sin\theta
	\label{eq: 14}
\end{equation}
Note this formally vanishes at $\theta = \pi$, but it requires an infinite amount of squeezing to get there!

 \begin{figure}
\includegraphics[width=8cm]{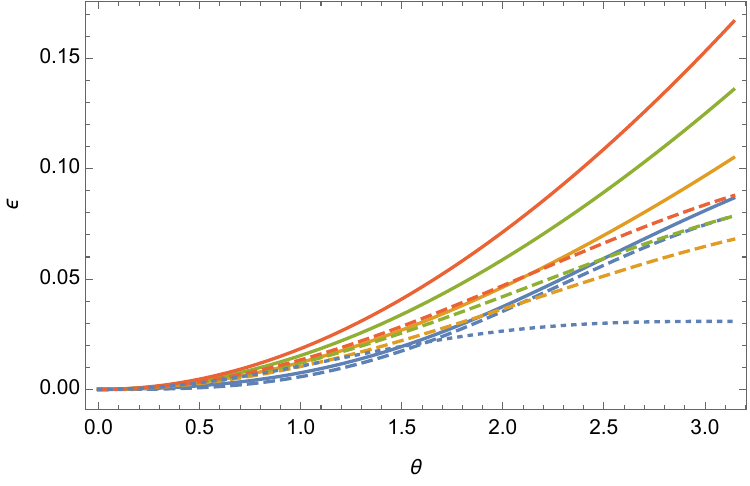}
\caption{Solid curves: error for an $x$-cat atomic state with (from bottom to top) $N=2,3,4$ and $5$ atoms and a coherent state with $\alpha^2 = 20 N$ photons.  Dashed curves: the corresponding error when an optimally squeezed field state is used instead.  Dotted curve: the maximum error (\ref{eq: 14}) for one atom, with a 20-photon coherent state. }
\end{figure}

Figure 3 summarizes graphically the main results from this Section.  The solid curves show the error for an $x$-cat atomic state and a coherent field state with $\alpha^2 = 20 N$ photons.  Because of this scaling the error increases roughly as $N \sim N^2/\bar n$.  The dashed curves show the error when an optimally-squeezed field state is used instead, giving the more favorable scaling as $\sqrt N \sim N\sqrt N/\bar n$.   

For reference, Figure 1 also shows the worst-case error for a single atom, as given by Eq.~(\ref{e17}), for a $20$-photon coherent state.  Note that, with the scaling used for the number of photons, the $N=2$ curves actually drop below the $N=1$ result for small values of $\theta$.  However, for large enough $\theta$, the error increase for $N=2$ eventually becomes quite noticeable, even with twice the number of photons in the field as for the $N=1$ case.  

Finally, we should note that we have not, strictly speaking, proved that the $x$-cat state gives the worst-case error for $N>1$.  In fact, the numerical simulations in the next Section show that this is not the case, but they also show that it is very hard to do significantly worse.

\section{Average over all initial states}

In this Section we consider the average error for an arbitrary single qubit rotation on a collective group of atoms, defined as the error averaged over all possible initial states. For $N$ atoms, we will write the initial state as
\begin{equation}
\ket{\psi_0} =\sum_{j=j_\text{min}}^{j_\text{max}}\sum_{k=1}^{k_\text{max}(j)} \sum_{m=-j}^j  C_{j,k,m}\ket{j,k,m}
\label{e20}
\end{equation}
where $j_\text{max} = N/2$ is the maximum pseudo-angular momentum one can form with $N$ two-level systems, and $j_\text{min}$ the corresponding minimum ($0$ for $N$ even and $1/2$ for $N$ odd).  The index $k$ is used to distinguish degenerate subspaces with the same value of $j$.  The corresponding degeneracy is given by $k_\text{max}$ \cite{tavisc}:
\begin{equation}
k_\text{max}=\frac{N!(2j+1)}{(\frac{N}{2}+j+1)!(\frac{N}{2}-j)!}
\label{e21}
\end{equation}

We will use an overbar to denote averages of the initial-state coefficients over the Haar measure. Trivial results are that $\overline{C^*_iC_j}=0$ (because of the random phases of the coefficients), and $\overline{C^*_iC_i} = 1/2^N$ (because all the basis states are equally probable).  For more complicated averages, we use the result that the Haar measure on $U(N)$ induces the uniform measure on the $N-1$ dimensional simplex for the vector formed by the squared coefficients $|C_i|^2$ \cite{zycz}.  This readily yields the basic results
\begin{equation}
\overline{|C_i|^4} = \frac{2}{2^N(2^N+1)}
\label{e22}
\end{equation}
and 
\begin{equation}
\overline{|C_i|^2 |C_j|^2} = \frac{1}{2^N(2^N+1)} \qquad (j\ne 1)
\label{e23}
\end{equation}

Just as in Section III, we find that only $J_+J_-$, $J_-J_+$, and $J_zJ_z$  contribute to the last term in (\ref{eq: 3}), and we obtain for this term (see Appendix for details)),
\begin{equation}
\overline{\av{\Psi_0|\Psi^{(2)}}}= -\frac{N}{32\alpha^2}\left(\theta^2 e^{-2r}+4\sin^2\frac{\theta}{2} e^{2r}\right)
	\label{eq: 15}
 \end{equation}
Unlike before, however, the $\overline{\|\av{\psi_0|\Psi^{(1)}}\|^2}$ term does not vanish, since for an arbitrary initial state we cannot expect  $\av{J_z}^2, \av{J_x}^2$ and $\av{J_y}^2$ to be zero, and thus, neither will their ensemble averages.  We can, however, expect the latter to be all equal to each other, because of the rotational symmetry of the Haar ensemble, which results in 
\begin{equation}
\begin{split}
\overline{\|\av{\psi_0|\Psi^{(1)}}\|^2}= \frac{\overline{\langle J_z\rangle^2}}{4\alpha^2\hbar^2}\left(\theta^2 e^{-2r}+4\sin^2\frac{\theta}{2} e^{2r} \right)
	\label{eq: 16}
 \end{split}
\end{equation}
At this point, $\overline{\langle J_z \rangle ^2}$ must be calculated. Again the details may be found in the Appendix.  The result is
\begin{equation}
\begin{split}
\overline{\langle J_z \rangle ^2} &= \frac{2\hbar^2}{2^N(2^N+1)}\sum_m^{N/2}m^2\binom{N}{\frac{N}{2}+m}\\&=\frac{N}{4(2^N+1)}\hbar^2
	\label{eq: 17}
 \end{split}
\end{equation}
where the sum begins at $m=0$ or $m=\frac{1}{2}$ if $N$ is even or odd, respectively. 

Putting it all together, the error average over initial states, or average error, is 
\begin{equation}
\overline{\epsilon} =\frac{1}{1+2^{-N}}\frac{N}{16\alpha^2}\left(\theta^2 e^{-2r}+4\sin^2\frac{\theta}{2} e^{2r}\right)
	\label{eq: 19}
 \end{equation}
In this expression, the contribution from (\ref{eq: 17}) is  the $2^{-N}$ term, which quickly becomes negligible for large values of $N$.  Except for this, and a factor of $16/\alpha^2$, the form of (\ref{eq: 19}) is the same as Eq.~(\ref{eq: 11}), obtained in Section III for two atoms in a cat state; hence, the optimal squeezing parameter will be the same, namely, Eq.~(\ref{eq: 12}):
\begin{equation}
r_\text{opt}=\frac{1}{4}\ln\left(\frac{\theta^2}{4 \sin^2(\theta/2)}\right)
	\label{e28}
\end{equation}
Note that, unlike the results obtained in Section III, this is independent of the number of atoms $N$.  It is also a relatively mild squeezing, as Figure 4 shows. 

\begin{figure}
\includegraphics[width=8cm]{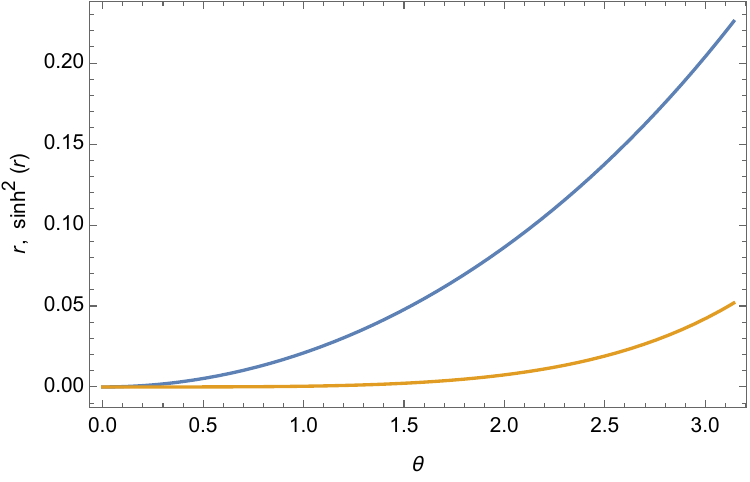}
\caption{Upper curve: The average ideal squeezing parameter~$r$ [Eq.~(\ref{e28})] to optimize the average error for any number of atoms, as a function of $\theta$.  Lower curve: the corresponding average number of photons ($\sinh^2 r$) added to the field as a result of the squeezing.}
\end{figure}

The corresponding optimum average error becomes 
\begin{equation}
\overline\epsilon_\text{opt}=\frac{1}{1+2^{-N}}\frac{N}{4\alpha^2}\theta\sin\frac \theta 2
\label{e29}
\end{equation}
(cf. Eq.~(\ref{eq: 13})). For large $N$ this, just like the coherent-state result (\ref{eq: 19}), scales as $N$, a much more favorable result than the worst-case scenarios considered in Section III.  This essentially means that the measure of the set of states leading to such large errors must become negligible for large $N$.  

\begin{figure}
\includegraphics[width=8cm]{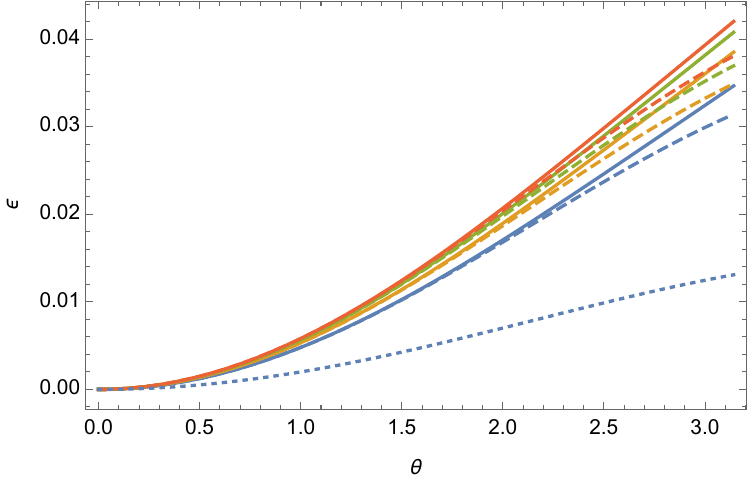}
\caption{{Solid curves: average error for a random initial state with (from bottom to top) $N=2,3,4$ and $5$ atoms and a coherent state with $\alpha^2 = 20 N$ photons.  Dashed curves: the corresponding error when an optimally squeezed field state is used instead.  Dotted curve: the maximum error for one atom, with a 20-photon coherent state. }}
\end{figure}

The average optimal error is bounded by the worst case scenario, which corresponds to $\theta=\pi$. This produces the inequality:
\begin{equation}
\begin{split}
\overline{\epsilon}_\text{opt} <\frac{1}{1+2^{-N}}\frac{\pi N}{4\bar{n}} 
	\label{eq: 22}
 \end{split}
\end{equation}

We find that, unlike in Section 2, the results presented above (Eqs.~(\ref{eq: 19}), (\ref{e28}) and (\ref{e29})) apply equally well to the $N=1$ and $N=2$ cases. We only note that the $2^{-N}$ term plays a most significant role in those cases.  This is apparent from Figure 5, which summarizes the results of this section.  As the $2^{-N}$ becomes negligible, the $N$ scaling takes over, and if the number of photons is scaled as $N$ (as in the figure), the curves begin to fall on top of each other; note that, when the $N$ scaling holds, the average error \emph{per atom} ($\epsilon/N$), for constant $\bar n$, does not increase with $N$.  Also, a comparison with Fig. 3 shows that the squeezing (\ref{e28}) does not make nearly as big a difference in the average error as the squeezing $r_{opt,x}$ (Eq.~(\ref{eq: 8})) did for the worst-case scenarios considered in Section III.

\begin{figure}
\includegraphics[width=8cm]{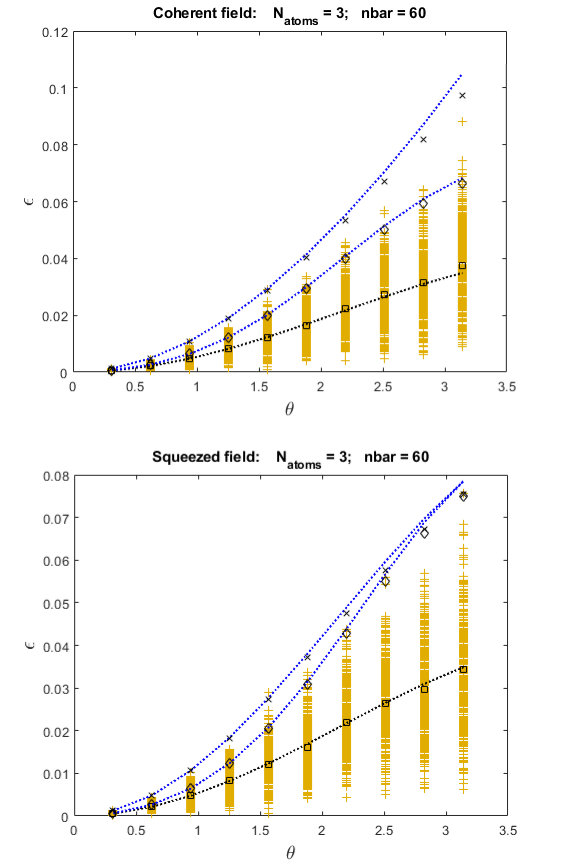}
\caption{{Numerical results (plus signs) of sampling $400$ initial random states of a system of 3 atoms, interacting with a field with $\alpha^2 =60$ in a coherent state (top figure) or optimally squeezed (bottom figure).  Dotted lines are theoretical results; crosses (resp. diamonds) are numerically calculated results for initial $x$-cat (resp. $z$-cat) states.  The lower dotted line is the theoretical average error, and the squares the numerically calculated average.}}
\end{figure}

To get a more complete picture of the physical situation, we have carried out numerical calculations with Haar-random ensembles of initial atomic states.  These are shown in Figures 6 and 7 for $N=3$ and $N=4$, respectively.  Several interesting features emerge.  First, the squeezing (\ref{e28}), as noted above, does not change the average error very much, but it does bring the $x$-cat state error down, even as it increases the error in the $z$-cat states; this is to be expected, since it is meant to reduce the amplitude fluctuations at the expense of increasing the phase fluctuations.  Second, the deviation of the theoretical curves from the numerically calculated results (most visible for the $x$-cat states in the region near $\theta = \pi$) is an indication of the limitations of perturbation theory for large values of the error $\epsilon$; increasing the number of photons in the field would result in smaller $\epsilon$ and overall better agreement.  Third, and most intriguingly, there is one data point for $N=3$ and $\theta = \pi/2$ that shows a (very slightly) greater error than the $x$-cat state, even for a coherent-state field.  We have examined the corresponding initial state; it is somewhat close to an $x$-cat state, but does not otherwise have any particularly distinguishing features.  This same state lies clearly above the $x$-cat curve when a squeezed field is used, but of course we would not necessarily expect the $x$-cat states to still give the largest error in this case.

\begin{figure}
\includegraphics[width=8cm]{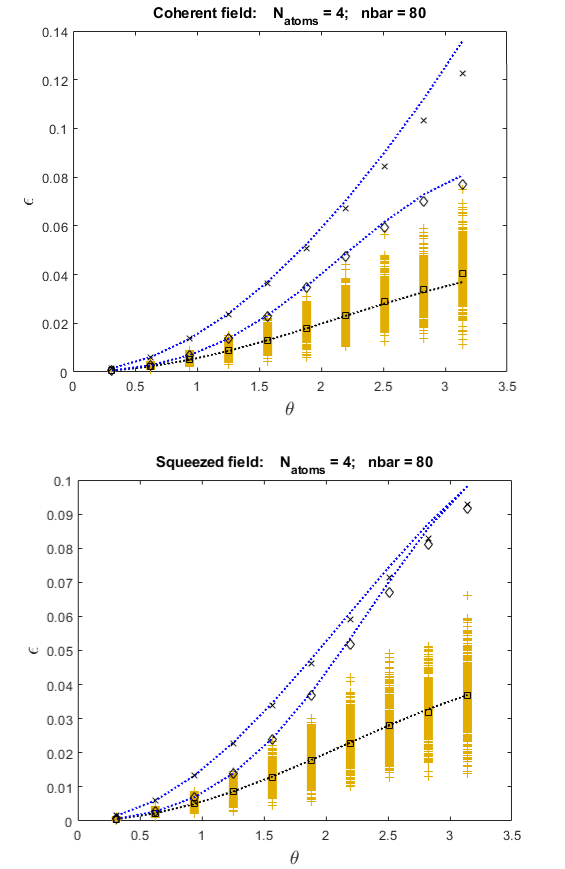}
\caption{{As in Figure 6, but for $N=4$, and $\alpha^2=80$.}}
\end{figure}

A comparison of Figures 6 and 7 shows that, as the number of atoms increases, the results of random sampling tend to cluster more closely around the average, and it becomes harder and harder to ``hit'' the states with the worst error scaling.  This makes sense: if there are states whose error scales as $N^2$, but the average error scales only as $N$, we should expect that the probability to find one of those states should decrease as $1/N$ as $N$ increases.

In any case, it is important to keep in mind that the squeezing (\ref{e28}), being independent of $N$, does not actually change the way the error scales for the cat states: for sufficiently large $N$, this would still grow as $N^2$.  Nevertheless, it does make a small difference for the relatively small values of $N$ shown in the figures.  In particular, since, in general, $x > \sin(x)$, Eq.~(\ref{e28}) will be positive, corresponding to a (mild) squeezing of the amplitude fluctuations at the expense of an increase in the phase fluctuations.  This will slightly reduce the error for the $x$-cat states and increase it for the $z$-cat states, as figures 6 and 7 show.



\section{Effects of adjusting the interaction time}

In this brief Section we wish to explore a possibility that was suggested to us by some recent numerical calculations, that, when the classical field is replaced by a quantized one, the optimum interaction time may not be exactly equal to the nominal, semiclassical one ($\theta/2g\alpha$).  If so, it might be possible to compensate in part for the error, by making a small adjustment $\Delta t$ to the interaction time.

It is relatively easy to modify our results to account for this possibility.  An obvious change is to replace $\theta$ by $\theta+\delta$ in all the formulas obtained for the $\ket{\Psi^{(1)}}$ and $\ket{\Psi^{(2)}}$ states, where $\delta = 2 g \alpha\Delta t$.  In addition, one has to consider that now the zero-th order state, which is the state rotated by the classical Hamiltonian, is not equal (in the second interaction picture) to the initial state, but has undergone an additional rotation by an angle $\theta$.  Hence, instead of (\ref{e4}), we must consider
\begin{equation}
\bra{\psi_0} e^{-i J_x \delta/\hbar}\ket{\Psi_0} + \av{\psi_0|\Psi^{(1)}} + \av{\psi_0|\Psi^{(2)}} 
\label{e31}
\end{equation}
This adds to all the error expressions a term 
\begin{equation}
1-\left|1-\frac{i\delta}{\hbar}\av{J_x}-\frac{\delta^2}{2\hbar^2}\av{J_x^2} \right|^2 \simeq \frac{\delta^2}{\hbar^2}\left(\av{J_x^2}-\av{J_x}^2\right)
\label{e32}
\end{equation}
(keeping only terms of order $\delta^2$).  For a $z$-cat state, and $N>2$, Eq.~(\ref{e32}) evaluates to $N\delta^2/4$, whereas for an $x$-cat state it gives $N^2 \delta^2/4$.  Taking the first case (since we would like to make the contribution of this term to the total error as small as possible), we get, from Eq.~(\ref{e10})
\begin{align}
\epsilon = &\frac{N\delta^2}{4} +  \frac{N}{16 \alpha^2}(\theta+\delta)^2e^{-2r} \cr
&+ \frac{N}{16 \alpha^2}\left[\sin^2(\theta+\delta) +4N\sin^4\left(\frac{\theta+\delta}{2}\right) \right]e^{2r}
\end{align}
Expanding this in powers of $\delta$ and optimizing, we get (assuming no squeezing, for simplicity)
\begin{equation}
\begin{split}
\delta_\text{$z$-cat, opt}=-\frac{\frac{1}{2}(N-1)\sin{2\theta}-N\sin\theta-\theta}{(N-1)\cos{2\theta}-N\cos\theta-1-4\bar{n}}
	\label{eq: 25}
 \end{split}
\end{equation}
This ends up typically being too small to make a substantial difference, because of the presence of the average number of photons term ($\bar n =|\alpha|^2$) in the denominator.  In addition, it tends to decrease further with increasing $N$, as can be seen by substituting $\theta=\pi$ in (\ref{eq: 25}).   

We will only consider explicitly the optimal case (from the point of view of maximizing the effect of $\delta$) of $N=2$ atoms in a cat state, where the error is given by
\begin{equation}
\epsilon(r,\delta)= \delta^2 + \frac{1}{4\alpha^2}\left((\theta+\delta)^2 e^{-2r}+4 \sin^4\left(\frac{\theta+\delta}{2}\right) e^{2r}\right)
	\label{e35}
 \end{equation}
 and the optimal $\delta$, again assuming zero squeezing, is
 \begin{equation}
\delta_\text{2, opt}=-\frac{\theta+\sin\theta(1-\cos\theta)}{1+4\bar{n}+\cos\theta-\cos(2\theta)}
	\label{eq: 26}
\end{equation}

\begin{figure}
\includegraphics[width=8cm]{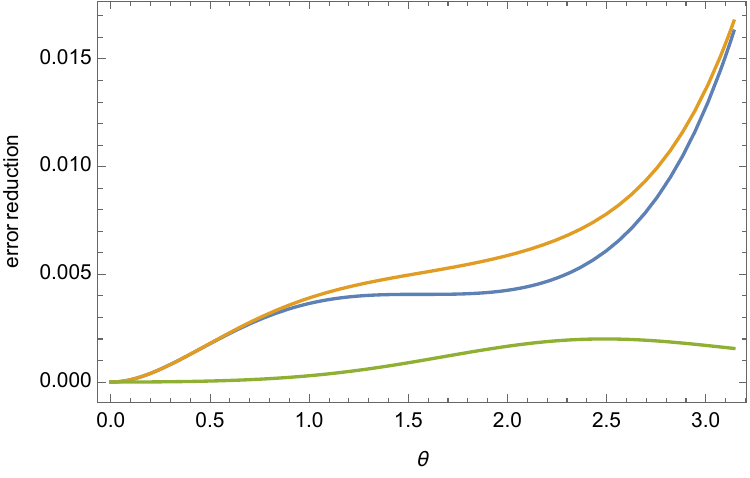}
\caption{{Absolute error reduction, $\epsilon(0,0)-\epsilon(r,\delta)$, for a rotation of a 2-atom cat state by a field with $\alpha = \sqrt{20}$, under three different strategies.  Lowest curve: $\delta = \delta_\text{2, opt}$ (from Eq.~(\ref{eq: 26}), $r=0$; middle curve, $\delta =0$, $r=r(\theta)$ from Eq.~(\ref{eq: 12}); top curve, $\delta = \delta_\text{2, opt}$ and  $r=r(\theta+\delta)$ (as in Eq.~(\ref{eq: 12}, but with $\theta$ replaced by $\theta+\delta$). }}
\end{figure}

Figure 8 compares three different strategies to reduce the error in a rotation of this state: ``underrotation'' by the angle given by (\ref{eq: 26}), squeezing the field by the amount given in (\ref{eq: 12}), and a combination of squeezing and underrotation.  We are assuming a very small number of photons in the field, $\alpha^2 = 20$, so the error (\ref{e35}) is relatively large: with both $r=0$ and $\delta = 0$, Eq.~(\ref{e35}) gives $\epsilon = 0.043$ for $\theta=\pi/2$ and $\epsilon = 0.173$ for $\theta = \pi$.  The figure shows the absolute reduction in the error resulting from each of the three strategies.  The maximum reduction amounts to about $3\%$ of the total error for $\theta=\pi/2$, and about $10\%$ for $\theta=\pi$.  While underrotation by itself is clearly inferior to squeezing,  there is a range of rotation angles where it may be a useful addition to it, at least for the very small number of photons considered here.

\section{Conclusions}
We have studied the error introduced by field quantization in the simultaneous performance of arbitrary rotations on a set of $N$ atoms.  Our results confirm the claim made in \cite{jgbozawa} that for some special, highly-entangled initial atomic states, and coherent-state fields, the error scales as $N^2$; surprisingly, though, the largest error comes from amplitude, rather than phase, fluctuations, for the ``cat'' states in the basis of $J_x = \hbar \sum_{i=1}^N \left(\ket e_i\bra g + \ket g_i \bra e \right)$.  

We have also shown, both numerically and analytically, that, despite the very unfavorable scaling exhibited by these special states, the error averaged over a random set of initial states scales only as the number of atoms, $N$, which means that the relative measure of the set of states where the error scales as $N^2$ decreases as $N$ increases.  Nevertheless, quantum computation schemes often make use of highly entangled states, sometimes of the ``cat'' form, and in some cases, such as error correction protocols, one may wish to manipulate several qubits simultaneously (perhaps of the order of 10 or more).  Thus, although all the errors calculated here decrease with the average number of photons in the field, as $1/\bar n$, and are not likely to be a problem in practice for even relatively low laser intensities, they still imply a lower bound on the power requirements for quantum computation.  

We have also studied error mitigation strategies, including adjustments to the interaction time (only useful for very low photon numbers and relatively large error rates), and, most importantly, squeezing of the driving field.  In particular, we have shown that, by using suitably squeezed field states, the error scaling for the most unfavorable initial atomic states can be reduced from $N^2$ to $N\sqrt N$, at the cost of an essentially negligible increase in the power required for the operation.  However, it is important to note that such squeezed states should not be used ``blindly,'' that is, in situations where one does not know much about the nature of the initial atomic state, since the squeezing that is optimal for an initial $x$-cat state will increase the error for a $z$-cat state (and vice-versa), and, moreover, for large $N$, either of them will \emph{worsen} the average error, by introducing large phase or amplitude field fluctuations, respectively.  This can be confirmed by, e.g., substituting either of the optimal ``cat'' squeezings from Eq.~(\ref{eq: 8}) in the equation for the average error, Eq.~(\ref{eq: 19}): the result scales, overall, as $N\sqrt N$, meaning that this scaling is now extended to the majority of initial states.

\begin{acknowledgements}
We acknowledge the MonArk NSF Quantum Foundry supported by the National Science Foundation Q-AMASE-i program under NSF award No. DMR-1906383.
\end{acknowledgements}

\appendix
\section{Some explicit mathematical results}

\subsection{1st and 2nd Order Perturbations}
The first- and second-order corrections to the joint atom-field state are given by
\begin{widetext}
\[\ket{\Psi^{(1)}}=\frac{-i}{2\hbar\alpha}\bigg[\theta J_x (\ap+{a^\prime}^\dagger)+i\sin\theta J_y(\ad-{a^\prime}^\dagger)+i(\cos\theta-1)J_z(\ap-{a^\prime}^\dagger)\bigg]\ket{\Psi_0}\]

\begin{equation*}
\begin{split}
\ket{\Psi^{(2)}}=\frac{-1}{4\hbar^2\alpha^2}\Bigg[\frac{\theta^2}{2}J_x^2(\ap+{a^\prime}^\dagger)^2-i(\cos\theta-1)J_xJ_y(\ap+{a^\prime}^\dagger)(\ap-{a^\prime}^\dagger)+i(\sin\theta-\theta)J_xJ_z(\ap+{a^\prime}^\dagger)(\ap-{a^\prime}^\dagger)\\+i(\theta\sin\theta+\cos\theta-1)J_yJ_x(\ap-{a^\prime}^\dagger)(\ap+{a^\prime}^\dagger)-\frac{\sin^2\theta}{2}J_y^2(\ap-{a^\prime}^\dagger)^2-\left(\frac{\theta}{2}+\frac{\sin2\theta}{4}-\sin\theta\right)J_yJ_z(\ap-{a^\prime}^\dagger)^2\\-i(\sin\theta-\theta\cos\theta)J_zJ_x(\ap-{a^\prime}^\dagger)(\ad+{a^\prime}^\dagger)+\left(\frac{\theta}{2}-\frac{\sin2\theta}{4}\right)J_zJ_y(\ap-{a^\prime}^\dagger)^2+\left(\frac{\sin^2\theta}{2}+\cos\theta-1\right)J_z^2(\ap-{a^\prime}^\dagger)^2\Bigg]\ket{\Psi_0}
\end{split}
\end{equation*}
\end{widetext}

%

\subsection{Derivation of $\overline{\langle J_z^2\rangle}$}
Since a random initial state must have spherical symmetry, we expect $\overline{\av{J_x J_y}} = \overline{\av{J_y J_z}} = \overline{\av{J_z J_x}} = 0$, because the expectation values involved are just as likely to be negative as positive.  Also by symmetry, we clearly must have $\overline{\av{J_x}^2}=\overline{\av{J_y}^2} = \overline{\av{J_z}^2}$, and since
\[\overline{\langle J_x^2 \rangle}+\overline{\langle J_x^2 \rangle}+\overline{\langle J_z^2 \rangle}=\overline{\langle J^2 \rangle}\]
we have
\[\overline{\langle J_z^2 \rangle} = \frac{1}{3}\overline{\langle J^2 \rangle}\]
We define our initial state to be (Eq.~(\ref{e20}))
\begin{equation}
\ket{\psi_0} =\sum_{j=j_\text{min}}^{j_\text{max}}\sum_{k=1}^{k_\text{max}(j)} \sum_{m=-j}^j  C_{j,k,m}\ket{j,k,m}
\label{a1}
\end{equation}
with (Eq.~(\ref{e21}))
\begin{equation*}
k_\text{max}=\frac{N!(2j+1)}{(\frac{N}{2}+j+1)!(\frac{N}{2}-j)!}
\end{equation*}
Then, since, by normalization, $\overline{|C_{j,m,k}|^2} = 1/2^N$, the average expectation value of $J_z^2$ is 
\begin{align*}
&\hbar^2\sum_{j=j_\text{min}}^{j_\text{max}}\sum_{k=1}^{k_\text{max}(j)} \sum_{m=-j}^j  (j+1)\overline{|C_{j,m,k}|^2}= \cr
&\qquad \frac{\hbar^2}{2^N}\sum_{j=j_\text{min}}^{N/2} j(j+1)(2j+1)\frac{N!(2j+1)}{(\frac{N}{2}+j+1)!(\frac{N}{2}-j)!}
\end{align*}
with $j_\text{min}= 0$ (for $N$ even) or  $1/2$ if $N$ is odd.
This sum simplifies to

\[\overline{\langle J^2 \rangle}=\frac{3N}{4}\hbar^2\]

\subsection{Derivation of $\overline{\langle J_z\rangle^2}$}
Although, for a random initial state, $\av{J_z}$ is just as likely to be positive or negative, $\av{J_z}^2$ has to be nonnegative, and therefore $\overline{\av{J_z}^2}$ cannot vanish.  With the same initial state (\ref{a1}), but with the summation limits omitted for clarity, we have
\begin{equation}
\overline{\langle J_z\rangle^2}=\hbar^2\overline{\bigg( \sum_j\sum_k\sum_m |C_{j,k,m}|^2\bigg)^2}
\label{a2}
\end{equation}
This sum would give us $2^{2N}$ terms, but a key point is that most of these terms will cancel each other.  Consider, specifically, a partial sum of the form
\begin{equation}
\sum_{m=-j}^j m \overline{|C_{j^\prime,k^\prime,m^\prime}|^2 |C_{j,k,m}|^2|}
\end{equation}
This will be zero if $j\ne j^\prime$, or $k\ne k^\prime$, since all the ensemble averages of products of absolute value squares of coefficients with different indices have the same value, and thus the negative values of $m$ in the sum will cancel out with the positive ones.  The same cancellation will happen, for a given $m^\prime$, between all the terms for which $m\ne m^\prime, -m^\prime$.  The multiple sum in (\ref{a2}) therefore reduces to 
\begin{align}
 &2\hbar^2 \sum_j\sum_k\sum_m m^2 \left(\overline{|C_{j,m,k}|^4}-\overline{|C_{j,m,k}|^2|C_{j,-m,k}|^2}\right) \cr
 &=\frac{2\hbar^2}{2^N(2^N+1)}  \sum_j\sum_k\sum_m m^2
 \end{align}
 (using Eqs.~(\ref{e22}), (\ref{e23})).  The values of $m$ in this expression will range, overall, from $0$ or $1/2$ (for $N$ even or odd, respectively) to $N/2$, and it is easy to figure out that over the whole sum a given value $m$ will appear $\binom{N}{\frac{N}{2}+m}$ times.  Therefore,
\begin{equation*}
\overline{\langle J_z\rangle^2}=\frac{2\hbar^2}{2^N(2^N+1)}\sum_m^{N/2}m^2\binom{N}{\frac{N}{2}+m}\]
Beginning with N=3, this sum gives the sequence 3, 8, 20, 48, 112,... which is equivalent to $2^{N-3}N$. Simplifying the above then yields
\[ \overline{\langle J_z\rangle^2} =\frac{N}{4(2^N+1)}\hbar^2\]

\end{document}